\documentclass[%
aps,
prl,
reprint,
amsmath,amssymb,
groupedaddress,
floatfix,
showpacs,
superscriptaddress,
]{revtex4-2}

\usepackage[dvipdfmx]{graphicx,color}
\usepackage{dcolumn}
\usepackage{bm}
\usepackage[mathlines]{lineno}
\usepackage{ulem} 

\begin{document}


\title{Evidence for $\bm{X(3872)\rightarrow J/\psi\pi^+\pi^-}$ 
produced in single-tag two-photon interactions}

\noaffiliation
\affiliation{University of the Basque Country UPV/EHU, 48080 Bilbao}
\affiliation{University of Bonn, 53115 Bonn}
\affiliation{Brookhaven National Laboratory, Upton, New York 11973}
\affiliation{Budker Institute of Nuclear Physics SB RAS, Novosibirsk 630090}
\affiliation{Faculty of Mathematics and Physics, Charles University, 121 16 Prague}
\affiliation{Chonnam National University, Gwangju 61186}
\affiliation{University of Cincinnati, Cincinnati, Ohio 45221}
\affiliation{Deutsches Elektronen--Synchrotron, 22607 Hamburg}
\affiliation{University of Florida, Gainesville, Florida 32611}
\affiliation{Department of Physics, Fu Jen Catholic University, Taipei 24205}
\affiliation{Key Laboratory of Nuclear Physics and Ion-beam Application (MOE) and Institute of Modern Physics, Fudan University, Shanghai 200443}
\affiliation{Justus-Liebig-Universit\"at Gie\ss{}en, 35392 Gie\ss{}en}
\affiliation{Gifu University, Gifu 501-1193}
\affiliation{II. Physikalisches Institut, Georg-August-Universit\"at G\"ottingen, 37073 G\"ottingen}
\affiliation{SOKENDAI (The Graduate University for Advanced Studies), Hayama 240-0193}
\affiliation{Gyeongsang National University, Jinju 52828}
\affiliation{Department of Physics and Institute of Natural Sciences, Hanyang University, Seoul 04763}
\affiliation{University of Hawaii, Honolulu, Hawaii 96822}
\affiliation{High Energy Accelerator Research Organization (KEK), Tsukuba 305-0801}
\affiliation{J-PARC Branch, KEK Theory Center, High Energy Accelerator Research Organization (KEK), Tsukuba 305-0801}
\affiliation{Higher School of Economics (HSE), Moscow 101000}
\affiliation{Forschungszentrum J\"{u}lich, 52425 J\"{u}lich}
\affiliation{IKERBASQUE, Basque Foundation for Science, 48013 Bilbao}
\affiliation{Indian Institute of Science Education and Research Mohali, SAS Nagar, 140306}
\affiliation{Indian Institute of Technology Guwahati, Assam 781039}
\affiliation{Indian Institute of Technology Hyderabad, Telangana 502285}
\affiliation{Indian Institute of Technology Madras, Chennai 600036}
\affiliation{Indiana University, Bloomington, Indiana 47408}
\affiliation{Institute of High Energy Physics, Chinese Academy of Sciences, Beijing 100049}
\affiliation{Institute of High Energy Physics, Vienna 1050}
\affiliation{Institute for High Energy Physics, Protvino 142281}
\affiliation{INFN - Sezione di Napoli, 80126 Napoli}
\affiliation{INFN - Sezione di Torino, 10125 Torino}
\affiliation{J. Stefan Institute, 1000 Ljubljana}
\affiliation{Institut f\"ur Experimentelle Teilchenphysik, Karlsruher Institut f\"ur Technologie, 76131 Karlsruhe}
\affiliation{Kavli Institute for the Physics and Mathematics of the Universe (WPI), University of Tokyo, Kashiwa 277-8583}
\affiliation{Kennesaw State University, Kennesaw, Georgia 30144}
\affiliation{Department of Physics, Faculty of Science, King Abdulaziz University, Jeddah 21589}
\affiliation{Kitasato University, Sagamihara 252-0373}
\affiliation{Korea Institute of Science and Technology Information, Daejeon 34141}
\affiliation{Korea University, Seoul 02841}
\affiliation{Kyungpook National University, Daegu 41566}
\affiliation{P.N. Lebedev Physical Institute of the Russian Academy of Sciences, Moscow 119991}
\affiliation{Faculty of Mathematics and Physics, University of Ljubljana, 1000 Ljubljana}
\affiliation{Ludwig Maximilians University, 80539 Munich}
\affiliation{Luther College, Decorah, Iowa 52101}
\affiliation{Malaviya National Institute of Technology Jaipur, Jaipur 302017}
\affiliation{University of Maribor, 2000 Maribor}
\affiliation{Max-Planck-Institut f\"ur Physik, 80805 M\"unchen}
\affiliation{School of Physics, University of Melbourne, Victoria 3010}
\affiliation{University of Mississippi, University, Mississippi 38677}
\affiliation{University of Miyazaki, Miyazaki 889-2192}
\affiliation{Moscow Physical Engineering Institute, Moscow 115409}
\affiliation{Graduate School of Science, Nagoya University, Nagoya 464-8602}
\affiliation{Kobayashi-Maskawa Institute, Nagoya University, Nagoya 464-8602}
\affiliation{Universit\`{a} di Napoli Federico II, 80126 Napoli}
\affiliation{Nara Women's University, Nara 630-8506}
\affiliation{National Central University, Chung-li 32054}
\affiliation{National United University, Miao Li 36003}
\affiliation{Department of Physics, National Taiwan University, Taipei 10617}
\affiliation{H. Niewodniczanski Institute of Nuclear Physics, Krakow 31-342}
\affiliation{Nippon Dental University, Niigata 951-8580}
\affiliation{Niigata University, Niigata 950-2181}
\affiliation{Novosibirsk State University, Novosibirsk 630090}
\affiliation{Osaka City University, Osaka 558-8585}
\affiliation{Pacific Northwest National Laboratory, Richland, Washington 99352}
\affiliation{Panjab University, Chandigarh 160014}
\affiliation{Peking University, Beijing 100871}
\affiliation{University of Pittsburgh, Pittsburgh, Pennsylvania 15260}
\affiliation{Punjab Agricultural University, Ludhiana 141004}
\affiliation{Research Center for Nuclear Physics, Osaka University, Osaka 567-0047}
\affiliation{Department of Modern Physics and State Key Laboratory of Particle Detection and Electronics, University of Science and Technology of China, Hefei 230026}
\affiliation{Seoul National University, Seoul 08826}
\affiliation{Showa Pharmaceutical University, Tokyo 194-8543}
\affiliation{Soochow University, Suzhou 215006}
\affiliation{Soongsil University, Seoul 06978}
\affiliation{Sungkyunkwan University, Suwon 16419}
\affiliation{School of Physics, University of Sydney, New South Wales 2006}
\affiliation{Department of Physics, Faculty of Science, University of Tabuk, Tabuk 71451}
\affiliation{Tata Institute of Fundamental Research, Mumbai 400005}
\affiliation{Department of Physics, Technische Universit\"at M\"unchen, 85748 Garching}
\affiliation{School of Physics and Astronomy, Tel Aviv University, Tel Aviv 69978}
\affiliation{Toho University, Funabashi 274-8510}
\affiliation{Department of Physics, Tohoku University, Sendai 980-8578}
\affiliation{Earthquake Research Institute, University of Tokyo, Tokyo 113-0032}
\affiliation{Department of Physics, University of Tokyo, Tokyo 113-0033}
\affiliation{Tokyo Institute of Technology, Tokyo 152-8550}
\affiliation{Tokyo Metropolitan University, Tokyo 192-0397}
\affiliation{Utkal University, Bhubaneswar 751004}
\affiliation{Virginia Polytechnic Institute and State University, Blacksburg, Virginia 24061}
\affiliation{Wayne State University, Detroit, Michigan 48202}
\affiliation{Yamagata University, Yamagata 990-8560}
\affiliation{Yonsei University, Seoul 03722}
  \author{Y.~Teramoto}\affiliation{Osaka City University, Osaka 558-8585} 
  \author{S.~Uehara}\affiliation{High Energy Accelerator Research Organization (KEK), Tsukuba 305-0801}\affiliation{SOKENDAI (The Graduate University for Advanced Studies), Hayama 240-0193} 
  \author{M.~Masuda}\affiliation{Earthquake Research Institute, University of Tokyo, Tokyo 113-0032}\affiliation{Research Center for Nuclear Physics, Osaka University, Osaka 567-0047} 
  \author{I.~Adachi}\affiliation{High Energy Accelerator Research Organization (KEK), Tsukuba 305-0801}\affiliation{SOKENDAI (The Graduate University for Advanced Studies), Hayama 240-0193} 
  \author{H.~Aihara}\affiliation{Department of Physics, University of Tokyo, Tokyo 113-0033} 
  \author{S.~Al~Said}\affiliation{Department of Physics, Faculty of Science, University of Tabuk, Tabuk 71451}\affiliation{Department of Physics, Faculty of Science, King Abdulaziz University, Jeddah 21589} 
  \author{D.~M.~Asner}\affiliation{Brookhaven National Laboratory, Upton, New York 11973} 
  \author{H.~Atmacan}\affiliation{University of Cincinnati, Cincinnati, Ohio 45221} 
  \author{T.~Aushev}\affiliation{Higher School of Economics (HSE), Moscow 101000} 
  \author{R.~Ayad}\affiliation{Department of Physics, Faculty of Science, University of Tabuk, Tabuk 71451} 
% ...

\collaboration{The Belle Collaboration}


\begin{abstract}
We report the first evidence for $X(3872)$ production 
in two-photon interactions by tagging either the electron or the positron in 
the final state, exploring the highly virtual photon region. 
The search is performed in 
$e^+e^-\rightarrow e^+e^- J/\psi\pi^+\pi^-$, 
using 825~fb$^{-1}$ of data collected by the Belle detector operated 
at the KEKB $e^+e^-$ collider. 
We observe three $X(3872)$ candidates, where the expected background
is $0.11\pm 0.10$ events, with a significance of 3.2$\sigma$.
We obtain an estimated value for $\tilde{\Gamma}_{\gamma\gamma}
{\cal B}(X(3872)\rightarrow J/\psi\pi^+\pi^-)$ 
assuming the $Q^2$ dependence predicted by 
a $c\bar{c}$ meson model,
where $-Q^2$ is the invariant mass-squared of the virtual photon.
No $X(3915)\rightarrow J/\psi\pi^+\pi^-$ candidates are found.
\end{abstract}


\maketitle

The charmonium-like state $X$(3872) has been observed 
in various interactions since its first 
observation in $B\rightarrow K J/\psi \pi^+\pi^-$ decays~\cite{X3872-Belle}. 
Its spin, parity, and charge conjugation are determined to be 
$1^{++}$~\cite{X3872-LHCb}, but its internal structure is 
still a puzzle~\cite{LHCb-width-1,LHCb-width-2}. 
Subsequent to the spin-parity determination, the $X(3872)$ has
not been searched for in two-photon interactions because 
axial-vector particles are forbidden to decay to 
two real photons~\footnote {
The $X(3872)$ was searched for in two-photon interactions 
before its spin-parity
determination: S. Dobbs {\it et al.} (CLEO Collaboration), 
Phys. Rev. Lett. {\bf 94}, 032004 (2005).}. 
However, mesons with $J^{PC}=1^{++}$ can 
be produced if one or both photons 
are highly virtual~\cite{SBG}---denoted as $\gamma^*$. 

We perform the first search for a $1^{++}$ charmonium state in two-photon
interactions  
using $e^+e^-\rightarrow e^+e^-X(3872)$, 
where one of the final-state electrons, referred to as a 
tagging electron, 
is observed, and the other scatters at an extremely 
forward (backward) angle and is not detected~\footnote 
{We use {``electron"} to 
denote both electron and positron.}. 
Such events are called single-tag events. 
The $X$(3872) is reconstructed via its decay to $J/\psi\pi^+\pi^-$ 
($J/\psi\rightarrow\ell^+\ell^-$). 
By measuring the momentum of the tagging electron, 
we measure the $Q^2$ dependence of $X(3872)$ production, 
where $-Q^2$ is the invariant mass-squared of the virtual photon.
If the $X(3872)$ has a molecular component in its structure, it must 
have a steeper $Q^2$ dependence than the regular $c\bar{c}$ state.
Hence, the single-tag two-photon interactions provide information
on the structure of this state.
The value of the two-photon decay width, 
obtained from this
measurement, is sensitive to the internal structure of the $X(3872)$.
Early attempts to calculate such decay widths for
charmonium-like exotic states have been reported 
in Ref.~\cite{VDM}.
We also search for the $X(3915)$ in the same final 
state through the 
$G$-parity-violating $J/\psi\rho^0$ ($\rho^0\rightarrow\pi^+\pi^-$)
channel, as well as $J/\psi\omega$ ($\omega\rightarrow\pi^+\pi^-$) 
decay\cite{X3915-1}.

We use 825~fb$^{-1}$ of data collected by the
Belle detector operated at the KEKB $e^+e^-$ asymmetric 
collider \cite{KEKB,PTEP-2}. 
The data were taken at the $\Upsilon(nS)$ 
resonances ($n\le 5$) and nearby energies, 
9.43~GeV$< \sqrt{s} <$ 11.03~GeV. 

The Belle detector is a general-purpose magnetic 
spectrometer~\cite{Detector,PTEP-1}. 
Charged-particle momenta 
are measured by a silicon vertex detector and a cylindrical drift
chamber. Electron and charged-pion identification 
relies on a combination of the drift chamber, time-of-flight 
scintillation counters, aerogel Cherenkov counters, 
and an electromagnetic calorimeter made of CsI(Tl) crystals. 
Muon identification relies on 
resistive plate chambers 
in the iron return yoke.

For Monte Carlo (MC) simulations,  
used to set selection criteria and
derive the reconstruction efficiency, we use 
TREPSBSS~\cite{TREPSBSS-1,TREPSBSS-2} to generate single-tag 
$e^+e^-\rightarrow e^+e^-X(3872)$ events in which the
$X(3872)$ decays to $J/\psi\pi^+\pi^-$ and $J/\psi$ decays leptonically. 
For simulating radiative $J/\psi$ decays, 
we use PHOTOS~\cite{PHOTOS-1,PHOTOS-2}.
A GEANT3-based program simulates the detector response~\cite{Geant}. 

Since one final-state electron is 
undetected, we select events with exactly five charged tracks, 
each coming from the interaction point (IP) 
and having $p_{\rm T}>0.1$~GeV/$c$,
with two or more having 
$p_{\rm T}>0.4$~GeV/$c$, where 
$p_{\rm T}$ is the transverse momentum with respect 
to the $e^+$ direction.

$J/\psi$ candidates are reconstructed by their decays 
to $e^+e^-$ or $\mu^+\mu^-$. 
A charged track is identified as an electron 
if its electron likelihood ratio, ${\cal L}_{e}/({\cal L}_{e}
+{\cal L}_{\pi})$, is greater than 0.66 and as a muon if it is not
selected as an electron and if its muon likelihood ratio, 
${\cal L}_{\mu}/({\cal L}_{\mu}+{\cal L}_{\pi}+{\cal L}_{K})$, is greater
than 0.66; ${\cal L}_{x}$ is the likelihood for a particle to 
be of species $x$~\cite{eID,MuonID}.
We require the mass of the lepton pair to be in the 
range 3.047--3.147 GeV/$c^2$.
In the calculation of the invariant mass of 
an $e^+e^-$ pair, we include
the four-momenta of radiated photons, having energy less 
than 0.2~GeV and angle relative to
an electron direction of less than 0.04~rad.

The tagging electron must have an electron likelihood ratio 
greater than 0.95 or $E/p$ greater than 0.87, where $E$ is the energy 
measured by the electromagnetic calorimeter 
and $p$ is the momentum of the particle. 
We require that the tagging electron have momentum 
above 1~GeV/$c$ and $p_{\rm T}> 0.4$~GeV/$c$. 
The electron momentum includes the momenta of radiated photons, 
using the same requirements as for the electrons from $J/\psi$ decays.

We identify a charged track as a pion if it satisfies the likelihood
ratio criteria of ${\cal L}_{\pi}/({\cal L}_{\pi}+{\cal L}_{K}) > 0.2$, 
${\cal L}_{\mu}/({\cal L}_{\mu}+{\cal L}_{\pi}+{\cal L}_{K}) < 0.9$, 
${\cal L}_{e}/({\cal L}_{e}+{\cal L}_{\pi}) < 0.6$, and its $E/p$ is 
less than 0.8~\cite{PID}.
Events should have no photons with energy above 
0.4~GeV or $\pi^0$ candidates with $\chi^2$ from the  
mass-constrained fit less than 4.0.

As the $X(3872)$ should be back-to-back
with the tagging electron projected in the plane
perpendicular to the beam axis, we require
the difference between their azimuthal angles be 
in the range ($\pi\pm 0.1$) rad.

\begin{figure}[htb]
  \centering
  \includegraphics[width=0.50\textwidth]{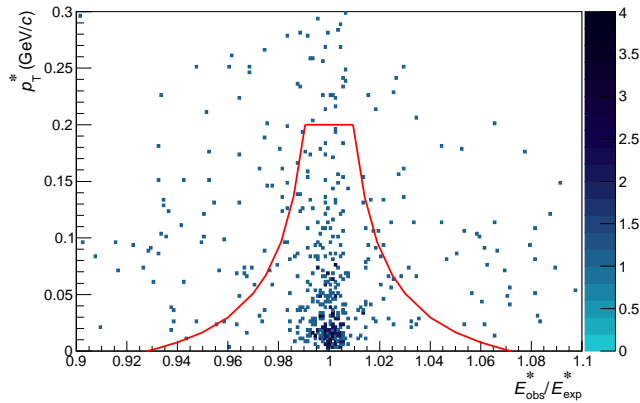}
  \caption{\label{fig:Hyper}
$p^{*}_{\rm T}$ vs.~$E^*_{\rm obs}/E^*_{\rm exp}$ 
distribution from data.
The (red) line shows the selection criteria applied
to $p^{*}_{\rm T}$ and 
$E^*_{\rm obs}/E^*_{\rm exp}$; 
events below the line are accepted.
}
\end{figure}

The total visible transverse momentum of the event,
$p^{*}_{\rm T}$~\footnote{
The $e^+e^-$ center-of-mass quantities are indicated by asterisks.}, 
should be less than 0.2~GeV/$c$.
We also require that the measured energy of 
the $J/\psi\pi^+\pi^-$ system, $E^*_{\rm obs}$, 
be consistent with the expectation, $E^*_{\rm exp}$, 
calculated from the momentum of the tagging
electron and the direction and invariant mass of the $J/\psi\pi^+\pi^-$
system, imposing energy-momentum conservation.
Since the energy and total transverse
momentum are correlated, we impose a two-dimensional 
criterion
\begin{equation}
{\displaystyle 
(p^{*}_{\rm T}+40~{\rm MeV/}c)
\left(\frac{|E^*_{\rm obs}-E^*_{\rm exp}|}{E^*_{\rm exp}}
+0.003\right)< 3~{\rm MeV/}c.
}
\end{equation}
Figure~\ref{fig:Hyper} shows the distribution of events
and these selection criteria in the 
$p^{*}_{\rm T}$ vs. $E^*_{\rm obs}/E^*_{\rm exp}$ plane.

Finally, we place a requirement on the missing 
momentum of the event, equal to the momentum of
the unmeasured electron that goes down the beam pipe. 
We require the missing-momentum projection 
in the $e^-$ beam direction in the 
center-of-mass frame be less than $-0.4$~GeV/$c$ for 
$e^-$-tagging events and greater than 0.4~GeV/$c$ for 
$e^+$-tagging events.

We search for $X(3872)$ and $X(3915)$ 
by looking for events in the $J/\psi\pi^+\pi^-$ 
mass distribution, $M(J/\psi\pi^+\pi^-)$. 
The reconstructed mass resolution is
expected to be 2.5~MeV/$c^2$ from the MC simulation.
We define two signal regions: 3.867--3.877~GeV/$c^2$ 
for the $X(3872)$ and 3.895--3.935~GeV/$c^2$ for the $X(3915)$. 
The former accommodates the $X(3872)$ with a known mass of 
$3871.69\pm 0.17$~MeV/$c^2$ and a 
decay width less than $1.2$~MeV~\footnote{
Recent measurements of the decay width show  
$\Gamma^{\rm BW}_{X(3872)}=0.96^{+0.19}_{-0.18}\pm 
0.21$~MeV~\cite{LHCb-width-1} and $\Gamma^{\rm BW}_{X(3872)}=1.39
\pm 0.24 \pm 0.10$~MeV~\cite{LHCb-width-2}.};
the latter accommodates the $X(3915)$ with a known mass of  
$3918.4\pm 1.9$~MeV/$c^2$ and a decay width of $20\pm 5$~MeV.
We constrain the $J/\psi$ mass to 3.09690~GeV/$c^2$ 
when we calculate 
$M(J/\psi\pi^+\pi^-)$~\cite{PDG}. 

\begin{figure}[htb]
  \centering
  \includegraphics*[width=0.5\textwidth]{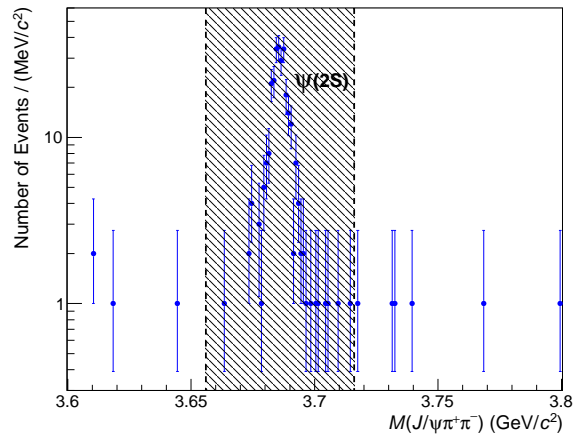}
    \caption{\label{fig:psi2S}$M(J/\psi\pi^+\pi^-)$ distribution 
shown with the $\psi(2S)$ veto (shaded gray region).
  }
\end{figure}

\begin{figure}[htb]
  \centering
  \includegraphics*[width=0.5\textwidth]{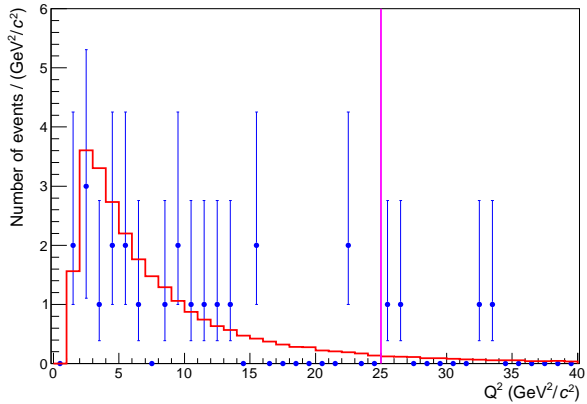}
    \caption{\label{fig:Q2_both}$Q^2$ distribution for data
(blue dots) and MC (red histogram). 
The area of MC distribution is normalized to that of data.
The vertical (magenta) line indicates the
selection requirement.
  }
\end{figure}

The dominant background, centered at 3.686~GeV/$c^2$, 
arises from radiatively produced $\psi(2S)$, 
$e^+e^-\rightarrow e^+e^-\psi(2S)$,
with $\psi(2S)\rightarrow J/\psi\pi^+\pi^-$. 
Figure~\ref{fig:psi2S} shows the
$M(J/\psi\pi^+\pi^-)$ distribution in data in the vicinity 
of $\psi(2S)$. Although the width of the $\psi(2S)$ peak 
is 2.7~MeV/$c^2$, it has a tail on the higher mass side. 
This feature was also seen in previous
studies of $J/\psi\pi^+\pi^-$ produced by 
initial-state radiation~(ISR)~\cite{ISR}. 
To remove $\psi(2S)$ events, we veto events within 0.03~GeV/$c^2$ 
of the $\psi(2S)$ mass, 3.686~GeV/$c^2$. 
Figure~\ref{fig:Q2_both} shows the $Q^2$ distribution 
after removing those events, where 
$Q^2 = 2(p_{\rm in}\cdot p_{\rm out}-m_{e}^2 c^2)$ 
and $p_{\rm in}$ and $p_{\rm out}$ are the 
four-momenta of the incoming~(beam) and outgoing~(tagging) 
electrons and $m_e$ is the electron mass.
In Fig.~\ref{fig:Q2_both}, data are dominated by background
events while MC is pure $X(3872)$.
Since two-photon processes are strongly suppressed at high
$Q^2$, we require $Q^2 < 25$~GeV$^2$/$c^2$ to 
reduce non-two-photon
background. 
Our measurement is insensitive for 
$Q^2 < 1.5$~GeV$^2$/$c^2$ due to low reconstruction efficiency.

\begin{figure}[htb]
  \centering
  \includegraphics*[width=0.50\textwidth]{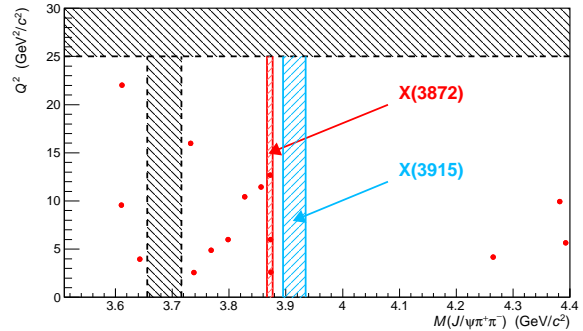}
    \caption{\label{fig:Open_Q2}Observed events (red dots) in the 
$Q^2$ vs.~$M(J/\psi\pi^+\pi^-)$ plane.
Three events are seen in the $X(3872)$ signal region 
(red lines with shade). 
The blue lines with shade show the $X(3915)$ signal region. 
The vetoed regions are shaded gray with dash lines.
  }
\end{figure}

Figure~\ref{fig:Open_Q2} shows the observed events in the $Q^2$ 
vs.~$M(J/\psi\pi^+\pi^-)$ plane. Three events are in 
the $X(3872)$ signal
region; no events are in the $X(3915)$ region.
The masses of the events in the $X(3872)$ signal region are 
3.8726, 3.8701 and 3.8742~GeV/$c^2$, 
averaging to  
3.8723$\pm$0.0012~GeV/$c^2$, where the uncertainty is statistical. 
At masses below the $X(3872)$ region, 3.716-3.867~GeV/$c^2$, 
there are six events, 
presumably from $\psi(2S)$ events; at masses above the $X(3872)$, there are
no events below 4.266~GeV/$c^2$, in region 
of the $Y(4260)$ mass.
A similar distribution was seen in the Belle ISR study~\cite{ISR}.
The 
$J/\psi\pi^+\pi^-$ events can also originate from
$t$-channel photon exchange with the emission of a virtual photon,
which we call internal bremsstrahlung (IB)~\cite{BREMS}.
Both processes produce $C$-odd $J/\psi\pi^+\pi^-$, like $\psi(2S)$, while 
the $C$-even $X(3872)$ peak can only be produced
from the two-photon process. 
The absence of a prominent $Y(4260)$ enhancement in our data
argues against non-negligible contribution from the $C$-odd
process through the decay 
$\gamma^* \rightarrow Y(4260)\rightarrow \gamma X(3872)$~\cite{BESIII}.
To estimate the background from IB, which has the same final-state 
particle configuration as our process and is hence difficult to separate, 
we use the ISR data~\cite{ISR}. 
By fitting the ISR data 
to our data in the region 3.5~GeV/$c^2$ $< M <$ 4.5~GeV/$c^2$, corrected
for the differences in the diagrams of $s$- and $t$-channels, we 
estimate the number of background events to be 
(3-5)$\times 10^{-2}$/(10~MeV/$c^2$) in the region between 3.8~GeV/$c^2$ 
and 4.2~GeV/$c^2$. 
This explains the absence of events between the $X(3872)$ and 4.26~GeV/$c^2$.

To estimate the background level in the $X(3872)$ signal region,
we fit a linear function 
\begin{equation}
{\rm max}(0, a(M(J/\psi\pi^+\pi^-)-3.872~{\rm GeV}/c^2)+b)
\end{equation}
to the data in the region $\pm 0.156$~GeV/$c^2$ centered 
at the $X(3872)$ mass, excluding the signal region; 
$a$ and $b$ are free in the fit. 
The width of $0.156$~GeV/$c^2$ is determined by the distance between 
the $X(3872)$ and the upper boundary, $3.716$~GeV/$c^2$, 
of the $\psi(2S)$ vetoed region. 
Using an unbinned extended maximum-likelihood fit, we obtain 
$a=-345\pm 195$~/(GeV/$c^2$)$^2$ and 
$b=10.5\pm 10.1$~/(GeV/$c^2$). 
This yields $n_{\rm b}=0.11\pm 0.10$ background events in the
$X(3872)$ signal window, where the uncertainty is statistical only. 

To derive the systematic uncertainty due to 
background modeling, we test 
two modified fitting functions. 
One is a power function, 
$a'/(M(J/\psi\pi^+\pi^-) - b')^{c'}$ with $b'$ set to 
2.4~GeV/$c^2$; the fit is insensitive to the value of $b'$.
This gives $n_{\rm b} = 0.096\pm 0.068$.
The other is a linear function with a break at 3.800~GeV/$c^2$,
$a''(M(J\psi\pi^+\pi^-) - 3.800~{\rm GeV}/c^2)+b''$ for 
$M(J\psi\pi^+\pi^-)<$3.800~GeV/$c^2$ and $b''$ for 
$M(J\psi\pi^+\pi^-)\ge$3.800~GeV/$c^2$, 
based on the shapes of the $M(J/\psi\pi^+\pi^-)$ distributions
in the ISR~\cite{ISR,ISR2} and the $e^+e^-$
annihilation studies~\cite{BESIII2,CLEO2}. 
This gives $n_{\rm b}=0.122\pm 0.095$. 
From the variations of $n_{\rm b}$ in the three forms, 
we derive $\pm 0.013$ for the systematic uncertainty.
This is negligible compared to the statistical uncertainty.
The estimated number of background events 
is $0.11\pm 0.10$, including statistical and systematic uncertainties.

With this background, the significance of three events is 3.2$\sigma$. 
For the $X(3872)$ signal, with three observed and 0.11 
expected background events, we calculate the number of signal events, 
$N_{\rm sig} = 2.9 {+2.2\atop -2.0}(\text{stat.}) \pm 0.1(\text{syst.})$, 
at 68\% confidence level~(C.L.).
For the $X(3915)$ signal, with zero observed and 0.3 expected
background events, we obtain $N_{\rm sig}< 2.14$ 
at 90\% C.L. The Feldman-Cousins method is used in 
both cases~\cite{FC}.

The differential cross section for the production of a resonance~($X$) 
in a single-tag two-photon interaction is expressed as~\cite{Ks}
\begin{eqnarray}
\label{Dcross}
{\displaystyle
\frac{{\rm d}\sigma_{ee}(X)}{{\rm d}Q^2}
}
& = & {\displaystyle
4\pi^2\left(1+\frac{Q^2}{M^2}\right)\frac{2J+1}{M^2}
\Gamma_{\gamma^*\gamma}(Q^2)
} \nonumber
\\ 
& & {\displaystyle
\times
2\left.{\frac{{\rm d}^2 L_{\gamma^*\gamma}}
{{\rm d}W{\rm d}Q^2}}\right|_{W=M}, 
}
\end{eqnarray}
where $L_{\gamma^*\gamma}$ is the single-tag luminosity function, 
$M$ is the resonance mass, 
$-Q^2$ is the invariant mass squared of the virtual photon, 
$\Gamma_{\gamma^*\gamma}(Q^2)$ is the 
$\gamma^*\gamma$ decay width, 
$W$ is the invariant mass
of the $\gamma^*\gamma$ system, and $J$ is the resonance spin.
The factor of two comes from the existence of two production modes: $e^-\gamma^*$ 
and $e^+\gamma^*$ scattering.

For a $J{=}1$ resonance, spin-parity conservation forbids
production at $Q^2=0$. 
To remove the 
$Q^2$-dependence from $\Gamma_{\gamma^*\gamma}(Q^2)$, 
we use the reduced $\gamma\gamma$ decay width 
$\tilde{\Gamma}_{\gamma\gamma}$ defined as~\cite{SBG,Aihara}
\begin{equation}
\label{Gtilde}
{\displaystyle
\tilde{\Gamma}_{\gamma\gamma}\equiv\lim_{Q^2\rightarrow 0}
\frac{M^2}{Q^2}\Gamma_{\gamma^*\gamma}^{\rm LT}(Q^2), 
}
\end{equation}
using its $Q^2$ dependence near zero; 
$\Gamma_{\gamma^*\gamma}^{\rm LT}$ is 
the $\gamma^*\gamma$ decay width corresponding to
a formation of the resonance from a longitudinal (virtual) photon 
and a transverse (real) photon.
Substituting this expression into Eq.~(\ref{Dcross}), we obtain
\begin{equation}\label{Dcross2}
{\displaystyle
\frac{{\rm d}\sigma_{ee}(X)}{{\rm d}Q^2}
= 4\pi^2\frac{3}{M^2} 2\frac{Q^2}{M^2}
\epsilon\tilde{\Gamma}_{\gamma\gamma}
2\left.{\frac{{\rm d}^2 L_{\gamma^*\gamma}}
{{\rm d}W{\rm d}Q^2}}\right|_{W=M}
}
\end{equation}
for $Q^2\ll M^2$, where an extra factor of two comes from the difference 
in the number of spin degrees of freedom: the longitudinal 
component has one degree of freedom and the transverse component 
has two with unpolarized incident photons.
In Eq.~(\ref{Dcross2}), $\epsilon$ is the ratio 
$L^{\rm LT}/L^{\rm TT}$, where $L^{\rm LT}$
is the luminosity function for the production of one longitudinally
polarized photon and one transversely polarized photon and 
$L^{\rm TT}$ is that for two transversely polarized photons. 
Using the Schuler-Berends-Gulik (SBG) model~\cite{SBG}{\footnote {
As a validation of the SBG model at higher $Q^2$, 
Ref. \cite{Ks} provides measurements of single-tag to no-tag ratios
for the $\gamma\gamma$ decay widths for $\chi_{c0}$ and $\chi_{c2}$, 
which agree with the predictions of this model. 
} } 
for $q\bar{q}$-type axial-vector mesons, this can be extended to 
higher $Q^2$~\cite{Aihara}:
\begin{equation}\label{SBG_formula}
{\displaystyle
\frac{{\rm d}\sigma_{ee}(X)}{{\rm d}Q^2}
= \tilde{\Gamma}_{\gamma\gamma} F(M,Q^2,\epsilon)
\left.{\frac{{\rm d}^2 L_{\gamma^*\gamma}}
{{\rm d}W{\rm d}Q^2}}\right|_{W=M}, 
}
\end{equation}
where 
\begin{equation}\label{F_func}
{\displaystyle
F(M,Q^2,\epsilon) = \frac{48\pi^2}{M^2} 
\frac{\frac{Q^2}{2M^2}+\epsilon}{\left(
1+\frac{Q^2}{M^2}\right)^3}\frac{Q^2}{M^2}, 
}
\end{equation}
accounting for contributions from helicity 0 and 1.
The SBG model, based on $c\bar{c}$, is the only model available at present 
that can reliably extend Eq.~(\ref{Dcross2}) to the higher $Q^2$ region:
Eq.~(\ref{F_func}).

To relate 
the number of signal 
events and the decay width, $\tilde{\Gamma}_{\gamma\gamma}$, 
we use Eqs.~(\ref{SBG_formula}) and (\ref{F_func}) 
assuming the $X(3872)$ 
is a pure $c\bar{c}$ state~\cite{SBG}, 
\begin{equation}
\label{SBG_formula2}
\begin{array}{l}
{\displaystyle
N_{\rm sig}} =  
{\displaystyle
L_{\rm int}
{\cal B}(X\rightarrow J/\psi \pi^+\pi^-) 
{\cal B}(J/\psi\rightarrow \ell^+\ell^-)
} 
\\ 
\times
{\displaystyle
\tilde{\Gamma}_{\gamma\gamma}
\int_{Q_{\rm min}^2}^{Q_{\rm max}^2} {\rm d}Q^2 F(M,Q^2,\epsilon)
\varepsilon_{\rm eff}(Q^2)
\left.{\frac{{\rm d}^2 L_{\gamma^*\gamma}
}{{\rm d}W{\rm d}Q^2}}\right|_{W=M}, 
} 
\end{array}
\end{equation}
where $\varepsilon_{\rm eff}(Q^2)$ is the $Q^2$-dependent 
reconstruction
efficiency, $L_{\rm int}$ is the integrated luminosity, 
${\cal B}(X\rightarrow J/\psi\pi^+\pi^-)$ is the branching fraction
of the $X(3872)$ to $J/\psi\pi^+\pi^-$, and 
${\cal B}(J/\psi\rightarrow \ell^+\ell^-)
= 0.1193$ is the branching fraction of $J/\psi$ to lepton pairs~\cite{PDG}.
We estimate the reconstruction efficiency from MC, 
in which we model the
$X(3872)$ decay as $X(3872)\rightarrow J/\psi\rho^0$ with
$J/\psi\rightarrow \ell^+\ell^-$ and $\rho^0\rightarrow\pi^+\pi^-$
and with all daughter particles isotropically distributed
in the rest frames of their parents. 
The decay model via $\rho$ is motivated 
by the measured mass distributions~\cite{X3872-Belle,Belle-X,CDF-X}. 
It has a reconstruction efficiency 12\% higher 
than that for non-resonant $\pi^+\pi^-$; we include a 6\% systematic 
uncertainty to account for this. 
The angular distribution of the decay products of the 
$X(3872)$ negligibly
affects the reconstruction, as confirmed by simulating with 
an alternative model with decay angles of 
daughters from a $J^P = 1^+$ resonance with helicities 0 and 1.

Detection efficiencies range from 4\% to 8\%
for $Q^2$ between 3~GeV$^2/c^2$ and 25~GeV$^2/c^2$ and have smaller values for
$Q^2 < 3$~GeV$^2/c^2$.
They are estimated for our three center-of-mass energies
on the $\Upsilon(2S)$,
$\Upsilon(4S)$, and $\Upsilon(5S)$ resonances and average the
values weighted by their corresponding integrated luminosities.
We also average over the four detection modes given the two
tagging charges ($e^+$ and $e^-$) and the two $J/\psi$ decay modes 
($e^+e^-$ and $\mu^+\mu^-$).

The luminosity functions for our beam energies are 
calculated as functions of $Q^2$ using TREPSBSS.
We set $\epsilon=1$ as a convention for the present
application of Eq.~(\ref{F_func})\cite{SBG}.
After performing the $Q^2$ integration in Eq.~(\ref{SBG_formula2}), 
from $Q_{\rm min}^2 = 1.5$~GeV$^2$/$c^2$
to $Q_{\rm max}^2 = 25$~GeV$^2$/$c^2$, 
we obtain
\begin{equation}\label{Relation}
{\displaystyle
\tilde{\Gamma}_{\gamma\gamma} {\cal B}(X(3872)\rightarrow J/\psi\pi^+\pi^-) 
= (1.88\pm 0.24)~{\rm eV}\times N_{\rm sig}, 
}
\end{equation}
including the total systematic uncertainty from the integration.

The dominant systematic uncertainty on 
$\tilde{\Gamma}_{\gamma\gamma}{\cal B(}X\rightarrow J/\psi\pi^+\pi^-)$ 
is from the reconstruction efficiency, 
primarily due to differences between MC and data. 
The largest uncertainty, 7\%, is in the $J/\psi$ selection 
from the uncertainty of the $e^+e^-$ background level.
We estimate the total systematic uncertainty to be 13\%. 

From $N_{\rm sig}$, we determine
\begin{equation} \nonumber
\tilde{\Gamma}_{\gamma\gamma}{\cal B}(X\rightarrow J/\psi\pi^+\pi^-)
=  5.5^{+4.1}_{-3.8}~(\text{stat.}) 
\pm 0.7~(\text{syst.})~{\rm eV}. 
\end{equation}
To set 
a limit on $\tilde{\Gamma}_{\gamma\gamma}$, 
we need ${\cal B}(X\rightarrow J/\psi\pi^+\pi^-)$. 
We derive an upper limit, using the measured products 
of $B$-meson decay branching fractions and the $X(3872)$
decay branching fractions, 
${\cal B}(B^+\rightarrow K^+ X) {\cal B}(X\rightarrow
J/\psi\pi^+\pi^-~{\rm and~other~specific~final~states})$
\footnote{
From ${\cal B}(B^+\rightarrow K^+X){\cal B}(X\rightarrow
J/\psi\pi^+\pi^-) = (8.6\pm 0.6)\times 10^{-6}$ and
the sum over the measured products of the branching
fractions, 
${\cal B}(B^+\rightarrow K^+X){\cal B}(X\rightarrow 
J/\psi\pi^+\pi^-, J/\psi\gamma, \psi(2S)\gamma, D^0\bar{D^0}\pi^0,
\bar{D^{*0}}D^0)=(1.4\pm 0.4)\times 10^{-4}$, 
where we exclude $\bar{D}^{*0}\rightarrow \bar{D}^0\pi^0$, 
we obtain that ${\cal B}(X\rightarrow J/\psi\pi^+\pi^-) < 0.061$ 
using the Bayesian method at 90\% C.L. This limit 
is consistent with C.~Li and C.-Z.~Yuan, Phys.~Rev.~{\bf D 100},
094003 (2019).}.
With the measured lower limit~\cite{Belle-X,BaBaR-Jpp,PDG}, this gives 
$0.032<{\cal B}(X\rightarrow J/\psi\pi^+\pi^-)<0.061$ at 
90\% C.L.
Using the Feldman-Cousins method for three observed events and 0.11
background, we obtain $0.995<N_{\rm sig}<7.315$ at 90\% C.L.
This, with Eq.~(\ref{Relation}), divided by 
${\cal B}(X\rightarrow J/\psi\pi^+\pi^-)$, gives 
the $\tilde{\Gamma}_{\gamma\gamma}$ range: 20-500~eV.
This is consistent with values predicted for 
the $c\bar{c}$ model~\cite{SBG,VDM}. 
For a comparison of experimental results with 
non-$c\bar{c}$ models, we must wait for improved 
calculations in the future.

No events consistent with $X(3915)\rightarrow J/\psi \pi^+\pi^-$
are observed. 
This, combined with past measurements~\cite{X3915-1,X3915-2}, 
indicates no excess of $G$-parity-violating decays 
of $X(3915)$.

In summary, we find the first evidence 
for $X(3872)$ production in 
two-photon, $\gamma^*\gamma$, interactions. 
We observe three $X(3872)$ 
candidates with a significance of 3.2$\sigma$ and an estimated
yield of $2.9 {+2.2\atop -2.0}~(\text{stat.})~\pm 0.1~(\text{syst.})$. 
From this, we obtain $\tilde{\Gamma}_{\gamma\gamma}
{\cal B}(X(3872)\rightarrow J/\psi\pi^+\pi^-) = 5.5 {+4.1\atop -3.8}
(\text{stat.})\pm 0.7(\text{syst.})$~eV, 
assuming the $Q^2$ dependence 
of a $c\bar{c}$ meson model.
With future advances in calculations 
of $\tilde{\Gamma}_{\gamma\gamma}$ 
for non-$c\bar{c}$ states and higher luminosities accumulated by Belle II,
we expect this method will clarify our
understanding of the $X(3872)$. 

\vspace*{\baselineskip}

\begin{acknowledgments}
We are grateful to M.~Karliner for useful discussions. 
We thank the KEKB group for excellent operation of the
accelerator; the KEK cryogenics group for efficient solenoid
operations; and the KEK computer group, the NII, and 
PNNL/EMSL for valuable computing and SINET5 network support.  
We acknowledge support from MEXT, JSPS and Nagoya's TLPRC (Japan);
ARC (Australia); FWF (Austria); NSFC and CCEPP (China); 
MSMT (Czechia); CZF, DFG, EXC153, and VS (Germany);
DST (India); INFN (Italy); 
MOE, MSIP, NRF, RSRI, FLRFAS project, GSDC of KISTI and KREONET/GLORIAD (Korea);
MNiSW and NCN (Poland); MSHE, Agreement 14.W03.31.0026 (Russia); University of Tabuk (Saudi Arabia); ARRS (Slovenia);
IKERBASQUE (Spain); 
SNSF (Switzerland); MOE and MOST (Taiwan); and DOE and NSF (USA).
\end{acknowledgments}

\bibliography{ref}

\end{document}